\documentclass[sigconf,natbib=true]{acmart}
\AtBeginDocument{%
  \providecommand\BibTeX{{%
    \normalfont B\kern-0.5em{\scshape i\kern-0.25em b}\kern-0.8em\TeX}}}

\usepackage{float}

\usepackage{booktabs}
\usepackage{array}
\usepackage{balance} 
\usepackage{multirow}
\usepackage[normalem]{ulem}
\usepackage{color}
\definecolor{lightgray}{RGB}{215,215,215}
\definecolor{myred}{RGB}{210,109,91}
\usepackage{colortbl}  
\usepackage{xcolor}
\useunder{\uline}{\ul}{}
\usepackage{subfigure}
\usepackage{algorithm}  
\usepackage{algorithmicx}  
\usepackage[noend]{algpseudocode}  

\usepackage{amsmath}  
\usepackage{enumitem}
\usepackage{tabularx}
\usepackage[utf8]{inputenc}
\usepackage[english]{babel}
\usepackage{amsthm}
\usepackage{bm}
\newcommand{\ie}{\emph{i.e., }}
\newcommand{\eg}{\emph{e.g., }}

\newcommand{\cf}{\emph{cf. }}

\newlength\myindent
\floatname{algorithm}{Algorithm}

\copyrightyear{2025}
\acmYear{2025}
\setcopyright{cc}
\setcctype{by}
\acmConference[SIGIR '25]{Proceedings of the 48th International ACM SIGIR Conference on Research and Development in Information Retrieval}{July 13--18, 2025}{Padua, Italy}
\acmBooktitle{Proceedings of the 48th International ACM SIGIR Conference on Research and Development in Information Retrieval (SIGIR '25), July 13--18, 2025, Padua, Italy}\acmDOI{10.1145/3726302.3730053}
\acmISBN{979-8-4007-1592-1/2025/07}

\begin{document}

\title{Order-agnostic Identifier for Large Language Model-based Generative Recommendation}

\author{Xinyu Lin}
\email{xylin1028@gmail.com}
\affiliation{
\institution{National University of Singapore}
\city{}
\country{Singapore}
}

\author{Haihan Shi}
\email{shh924@mail.ustc.edu.cn}
\affiliation{
\institution{University of Science and Technology of China}
\city{Hefei}
\country{China}
}

\author{Wenjie Wang}
\email{wenjiewang96@gmail.com}
\authornote{Corresponding authors. This research is supported by A*STAR, CISCO Systems (USA) Pte. Ltd and National University of Singapore under its Cisco-NUS Accelerated Digital Economy Corporate Laboratory (Award I21001E0002), the National Research Foundation, Singapore under its National Large Language Models Funding Initiative, (AISG Award No: AISG-NMLP-2024-002). Any opinions, findings and conclusions or recommendations expressed in this material are those of the author(s) and do not reflect the views of National Research Foundation, Singapore.}
\affiliation{
\institution{University of Science and Technology of China}
\city{Hefei}
\country{China}
}

\author{Fuli Feng$^{*}$}
\email{fulifeng93@gmail.com}
\affiliation{
\institution{University of Science and Technology of China}
\city{Hefei}
\country{China}
}

\author{Qifan Wang}
\email{wqfcr@meta.com}
\affiliation{
\institution{Meta AI}
\city{Menlo Park}
\country{USA}
}

\author{See-Kiong Ng}
\email{seekiong@nus.edu.sg}
\affiliation{
\institution{National University of Singapore}
\city{}
\country{Singapore}
}

\author{Tat-Seng Chua}
\email{dcscts@nus.edu.sg}
\affiliation{
\institution{National University of Singapore}
\city{}
\country{Singapore}
}

\renewcommand{\shortauthors}{Xinyu Lin et al.}

\begin{abstract}
Leveraging Large Language Models (LLMs) for generative recommendation has attracted significant research interest, where item tokenization is a critical step. 
It involves assigning item identifiers for LLMs to encode user history and generate the next item. 
Existing approaches leverage either token-sequence identifiers, representing items as discrete token sequences, or single-token identifiers, using ID or semantic embeddings. Token-sequence identifiers face issues such as the local optima problem in beam search and low generation efficiency due to step-by-step generation.
In contrast, single-token identifiers fail to capture rich semantics or encode Collaborative Filtering (CF) information, resulting in suboptimal performance.


To address these issues, we propose two fundamental principles for item identifier design:
1) integrating both CF and semantic information to fully capture multi-dimensional item information, and
2) designing order-agnostic identifiers without token dependency, mitigating the local optima issue and achieving simultaneous generation for generation efficiency. 
Accordingly, we introduce a novel \textit{set identifier} paradigm for LLM-based generative recommendation, representing each item as a set of order-agnostic tokens. 
To implement this paradigm, we propose SETRec, which leverages CF and semantic tokenizers to obtain order-agnostic multi-dimensional tokens. 
To eliminate token dependency, SETRec uses a sparse attention mask for user history encoding and a query-guided generation mechanism for simultaneous token generation.
We instantiate SETRec on T5 and Qwen (from 1.5B to 7B). 
Extensive experiments on four datasets demonstrate its effectiveness across various scenarios (\eg full ranking, warm- and cold-start ranking, and various item popularity groups). 
Moreover, results validate SETRec's superior efficiency and show promising scalability on cold-start items as model sizes increase.

\end{abstract}

\ccsdesc[500]{Information systems~Recommender systems}
\keywords{Item Tokenization, Set Identifier, LLM-based Recommendation}

\maketitle

\section{Introduction}\label{sec:introduction}

\begin{figure}[t]
\setlength{\abovecaptionskip}{0.02cm}
\setlength{\belowcaptionskip}{-0.3cm}
\centering
\includegraphics[scale=1.1]{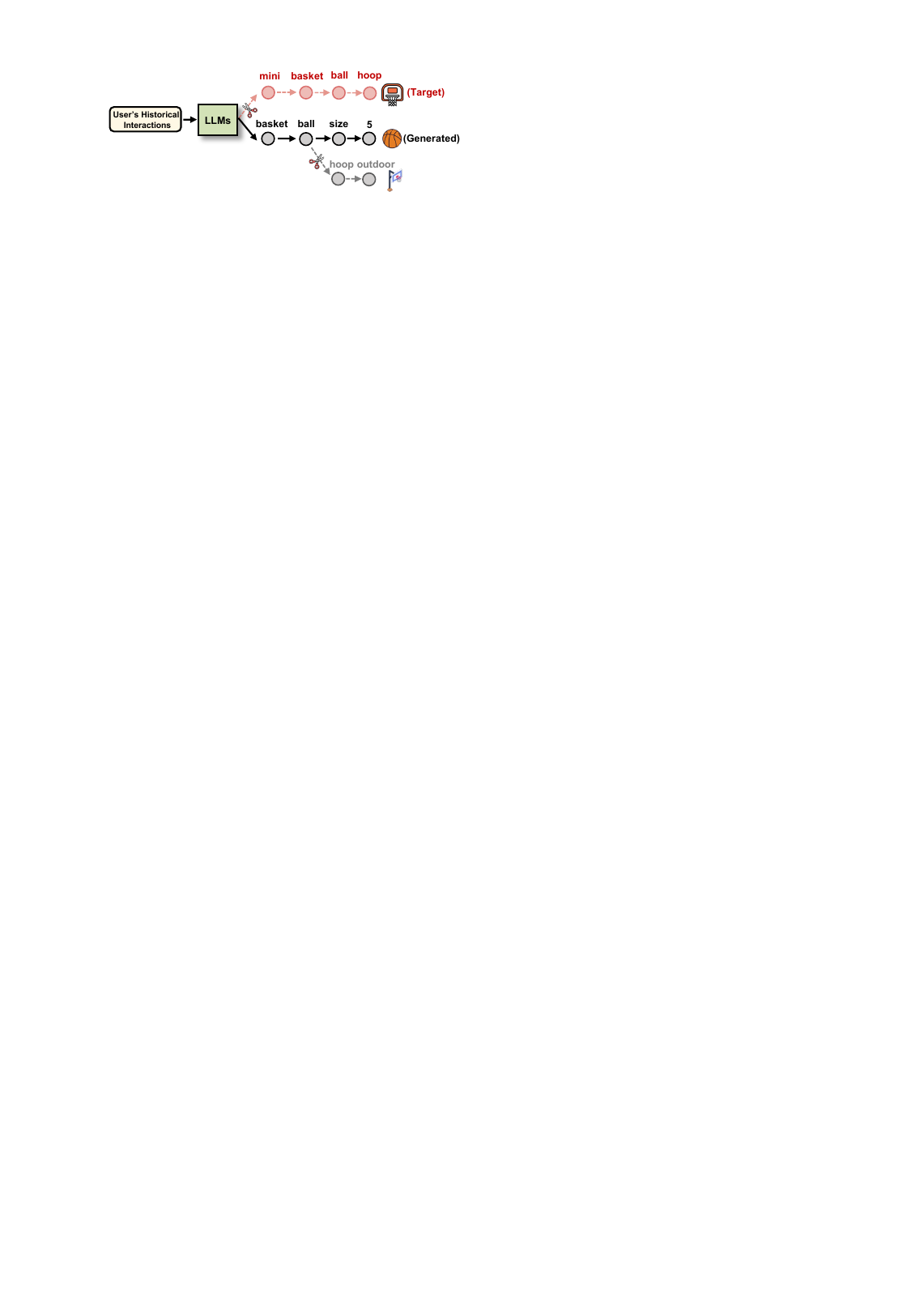}
\caption{An example of local optima issue in beam search in autoregressive item generation. The target item fails to be generated because the initial token has a low probability and hence is discarded at the early steps by beam search.}
\label{fig:local_optima}
\end{figure}

\begin{figure}[t]
\setlength{\abovecaptionskip}{0.02cm}
\setlength{\belowcaptionskip}{-0.3cm}
\centering
\includegraphics[scale=0.75]{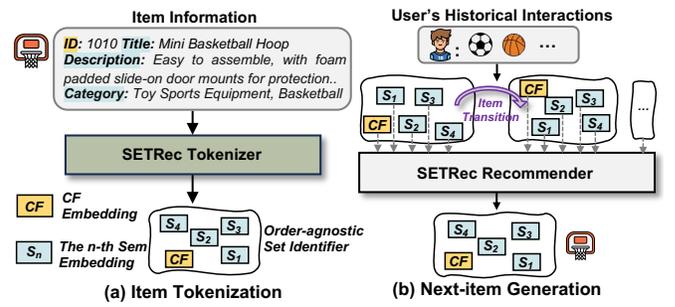}
\caption{Overview of SETRec. (a) Depiction of order-agnostic set identifiers representing items from multi-dimensional information. (b) SETRec emphasizes item sequential dependencies while removing token dependencies within items, which allows simultaneous generation to improve efficiency.}
\label{fig:intro}
\end{figure}

Large Language Models (LLMs) have recently demonstrated significant success in personalized recommendation, attracting widespread research interests~\cite{shi2024large,liao2024llara}.
Surpassing the traditional recommender models, LLMs excel in understanding complex user behaviors and diverse item characteristics due to their rich world knowledge and strong reasoning ability~\cite{touvron2023llama}. 
Typically, LLM-based recommenders transform the user's historical interactions into a token sequence to generate the target item as a recommendation. 
As shown in Figure~\ref{fig:intro}, a fundamental step of this process is item tokenization, which assigns each item an identifier to enable user history encoding and item generation. 
Therefore, item tokenization is essential in advancing LLM-based generative recommendation.

Existing item identifiers for LLM-based generative recommendation can be broadly categorized into two groups: 
\begin{itemize}[leftmargin=*]
    \item \textit{\textbf{Token-sequence identifier}} utilizes a discrete token sequence to represent multi-dimensional item information. 
    To generate items, LLMs use beam search to generate the top-$K$ item identifiers. 
    Despite the effectiveness, token-sequence identifiers suffer from the 
    1) \textit{local optima} issue~\cite{zhou2005beam} in the beam search. 
    As illustrated in Figure~\ref{fig:local_optima}, beam search greedily selects the sequence with top-$K$ probabilities at each generation step. 
    However, the initial tokens of the target identifier might not necessarily align with the user preference. 
    As such, the prefix of the target identifier has a low probability and will be pruned by beam search, causing inaccurate results. 
    2) \textit{Low generation efficiency} in the autoregressive generation, which requires multiple serial LLM calls, thereby causing unaffordable computing burdens~\cite{lin2024efficient} and severely hindering real-world deployments. 

    \item \textbf{\textit{Single-token identifier}} represents each item with a continuous token, \ie ID embedding or semantic embedding~\cite{liao2024llara,wang2024rethinking}. 
    To recommend items, LLMs first generate the next item embedding, which is then grounded to the item IDs with a linear projection layer, as exemplified by E4SRec~\cite{li2023e4srec} and LITE-LLM4Rec~\cite{wang2024rethinking}. 
    However, using single embedding often yields suboptimal performance. 
    Precisely, ID embeddings rely on sufficient interactions to capture Collaborative Filtering (CF) information, thus being vulnerable to long-tailed users/items.  
    Conversely, semantic embedding overlooks the modeling of CF information that is essential for personalized recommendations. 
\end{itemize}

Facing the above issues, a fundamental question arises: 
How can we design item identifiers to ensure effective and efficient LLM-based recommendations? 
Based on the above insights, we posit two principles. 
1) \textbf{Integration of semantic and CF information}. 
Semantic information can harness rich knowledge in LLMs to strengthen the generalization ability (\eg cold-start recommendation). 
Meanwhile, CF information leverages user behaviors to enrich the semantic modeling of user preference, enabling effective recommendations for users and items with rich interactions. 
2) \textbf{Order-agnostic Identifier.} 
Representing multi-dimensional item information (\eg semantic and CF information) with a single token might be suboptimal due to potential conflicts between different dimensions as proven in~\cite{wang2024learnable,zhang2022re4} (see empirical evidence in Section~\ref{sec:exp_hyper_param}). 
Therefore, it is necessary to utilize a set of tokens to effectively represent items with multi-dimensional information.
Nevertheless, 
multi-dimensional information is not necessarily dependent on each other (\eg ``price'' and ``category''). 
Moreover, ordered token sequences can risk the local optima issue. 
Hence, it is beneficial to disregard token dependencies in identifiers, which further facilitates simultaneous token generation, thus significantly improving inference efficiency. 


In this light, we introduce a novel paradigm of \textbf{\textit{set identifier}} for LLM-based generative recommendation.  
As shown in Figure~\ref{fig:intro}, it employs a set of order-agnostic tokens to represent each item with CF and semantic information. 
Nonetheless, it is non-trivial to eliminate token dependencies due to the following challenges: 
 
\begin{itemize}[leftmargin=*]
    \item For user history encoding, the transformed item sequence naturally introduces unnecessary token dependencies (\eg semantic tokens are dependent on the CF token), which might negatively affect user history encoding. 
    \item For the simultaneous generation of order-agnostic identifiers, 
    tokens are independently generated for each dimension (\eg CF). 
    This necessitates guidance on LLMs to generate tokens aligning well with each information dimension respectively. 
    \item 
    Since the tokens for different dimensions are generated independently, 
    the generated set identifier might be invalid items, which requires effective grounding to the existing items. 
\end{itemize}


To this end, we propose SETRec, an effective implementation of the set identifier paradigm. 
\textit{To integrate semantic and CF information}, SETRec leverages CF and semantic tokenizers to assign each item with an order-agnostic token set containing CF and semantic embeddings. 
\textit{To eliminate token dependencies}, 
1) for user history encoding, 
we propose a special sparse attention mask, which discards the visibility of other tokens within identifiers and retains access to previous identifiers. 
2) For simultaneous token generation, 
we introduce a query-guided generation mechanism, which adopts learnable vectors to guide LLMs to generate the embedding for each specific information dimension. 
3) To ground the generated embedding set to existing items, SETRec collects embeddings from all items as grounding heads to obtain the item scores for ranking. 
We instantiate SETRec on T5 and Qwen and evaluate it on four real-world datasets under various scenarios (\eg full ranking, warm- and cold-start ranking, and diverse item popularity groups) to demonstrate the effectiveness, efficiency, and generalization ability. 
Additionally, we evaluate SETRec on Qwen with different model sizes (\ie 1.5B, 3B, and 7B), exhibiting promising scalability on cold-start items as model size increases. 
The code and datasets are available at~\url{https://github.com/Linxyhaha/SETRec}.


The main contributions of this work are summarized as follows:
\begin{itemize}[leftmargin=*]
    \item We propose a novel set identifier paradigm for LLM-based generative recommendation, representing each item with a set of order-agnostic tokens integrating semantic and CF information. 
    \item 
    We propose SETRec to implement the novel paradigm, which introduces a query-guided generation mechanism with a sparse attention mask to achieve simultaneous generation without token dependencies, significantly boosting inference efficiency. 
    \item We instantiate SETRec on T5 and Qwen from 1.5B to 7B. Extensive experiments on four real-world datasets under various settings (\eg full ranking, warm- and cold-start ranking) validate its effectiveness, efficiency, generalization ability, and scalability. 
\end{itemize}
\section{Preliminaries}\label{sec:task_formulation}

\begin{figure}[t]
\setlength{\abovecaptionskip}{0.0cm}
\setlength{\belowcaptionskip}{-0.3cm}
\centering
\includegraphics[scale=0.4]{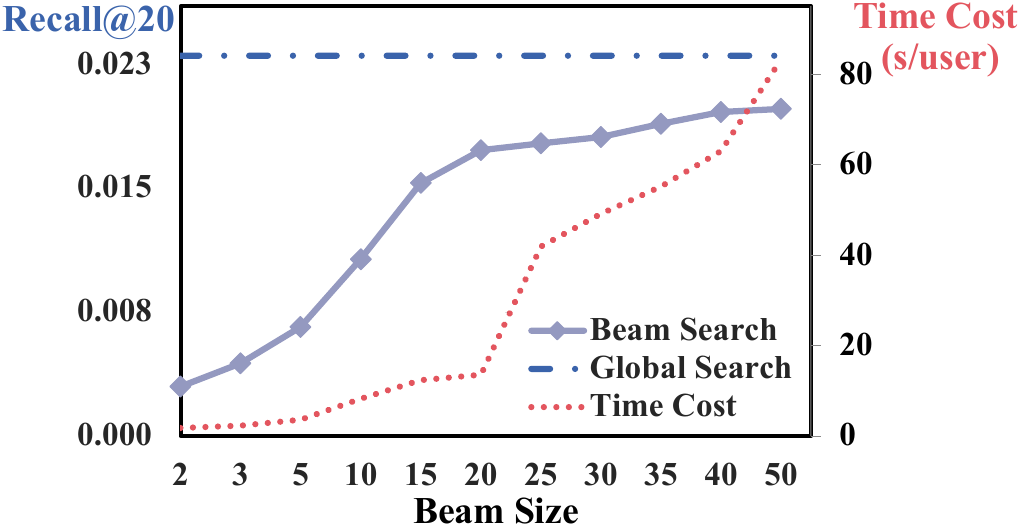}
\caption{Performance comparison between beam search and global search of LETTER on Toys. The global search is implemented by computing sequence probability for every item and ranking them based on the probabilities. }
\label{fig:beam_size}
\end{figure}

\vspace{2pt}
\noindent\textbf{LLM-based Generative Recommendation.} 
Harnessing LLMs' strong capabilities, LLM-based generative recommendation aims to use LLMs as recommenders to directly generate personalized recommendations. 
Formally, given the recommendation data 
$\mathcal{D}=\{\mathcal{S}_u|u\in\mathcal{U}, i\in\mathcal{I}\}$, where $\mathcal{S}_u = [i_1^{u}, i_2^{u}, \dots, i_L^{u}]$ is the user's historical interactions in chronological order and $L=|\mathcal{S}_u|$, 
the target is to utilize a tokenizer $f(\cdot)$ to tokenize items into item identifiers $\tilde{\mathcal{I}}$, 
and an LLM-based recommender model $\mathcal{M}(\cdot)$ to encode the transformed user history $\bm{x}= [f(i_{1}), f(i_{2}), \dots, f(i_{L})]$ and generate next item identifier. 


\vspace{2pt}
Bridging the language space and the item space, item identifier is a fundamental component for LLMs to encode user history and generate items. 
Existing identifiers can be divided into two groups:

\vspace{2pt}
\noindent$\bullet\quad$\textbf{\textit{Token-sequence identifier}} assigns each item with a discrete token sequence, \ie $\tilde{i} = [z_1, z_2, \dots, z_N]$, where $z_i$ is the discrete token. 
Given the user history $\mathcal{S}_u$, it is transformed to an identifier sequence $\bm{x}= [\tilde{i}_1, \tilde{i}_2, \dots, \tilde{i}_L]$, which is then encoded by LLMs to generate the next identifier $\hat{i}$ via autoregressive generation: 
\begin{equation}
\begin{aligned}
    &\hat{i}_t = \mathop{\arg\max}_{v\in\mathcal{V}} \mathcal{M}(v|\hat{i}_{<t},\bm{x}), \\
\end{aligned}
\end{equation} 
where $\hat{i}_t$ is the $t$-th token of identifier $\hat{i}$, $\hat{i}_{<t}$ represents the token sequence preceding $\hat{i}_t$, and 
$\mathcal{V}$ is the LLM vocabulary. 
Despite the effectiveness, generating token sequences would result in the local optima issue and inference inefficiency. 
As shown in Figure~\ref{fig:beam_size}, continuously increasing the beam size slightly improves recommendation accuracy, but remains inferior to globally optimal results. 
Worse still, the token-by-token generation requires multiple serial LLM calls, which significantly lowers the inference speed and hinders real-world applications.


\vspace{2pt}
\noindent$\bullet\quad$\textbf{\textit{Single-token identifier}} assigns each item with an ID or semantic embedding, \ie $\tilde{i}=\bm{z}$, 
which is usually obtained by a conventional CF recommender model (\eg SASRec~\cite{kang2018self})
or a pre-trained semantic extractor (\eg SentenceT5~\cite{ni2021sentence}). 
Given the transformed user history $\bm{x}= [f(i_{1}), f(i_{2}), \dots, f(i_{L})]$, it first generates the embedding:
\begin{equation}
\begin{aligned}
&\hat{i} = \text{LLM\_Layers}(\bm{x}), \\
\end{aligned}
\end{equation}
where $\text{LLM\_Layers}(\cdot)$ is the attention layers from the LLM $\mathcal{M}(\cdot)$. 
Based on the generated item embedding $\hat{i}$, 
an additional grounding head is added on top of the LLM layers to obtain the scores for all items for ranking. 
Although it improves inference efficiency by bypassing the token-by-token autoregressive generation, 
representing items with a single ID embedding struggles with items with fewer interactions while a single semantic embedding overlooks the crucial CF information, thus leading to suboptimal results. 

Based on the above insights, we
summarize two fundamental principles for identifier designs: 
1) integration of semantic and CF information, to leverage rich multi-dimensional item information, and  
2) order-agnostic identifier, to eliminate the unnecessary dependencies between tokens associated with an identifier, which can alleviate the local optima issue and improve generation efficiency.  
In this light, we introduce a novel set identifier paradigm, which employs a set of order-agnostic tokens to represent multi-dimensional item information. 

\section{SETRec}\label{sec:method}
\begin{figure}[t]
\setlength{\abovecaptionskip}{0.02cm}
\setlength{\belowcaptionskip}{-0.3cm}
\centering
\includegraphics[scale=0.88]{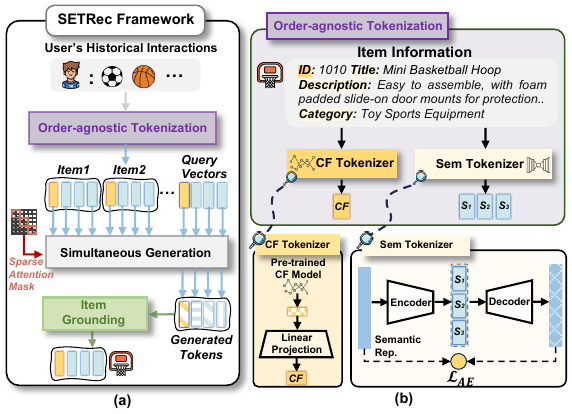}
\caption{(a) demonstrates SETRec framework, including order-agnostic item tokenization, and simultaneous item generation. The dependencies within identifiers and query vectors are eliminated by the sparse attention mask (see Figure~\ref{fig:sparse_attn} for details). (b) illustrates order-agnostic item tokenization via CF and semantic tokenizers.}
\label{fig:method_tokenizer}
\end{figure}

To implement the set identifier paradigm, we propose a framework called SETRec for effective and efficient LLM-based generative recommendation, including order-agnostic item tokenization and simultaneous item generation as illustrated in Figure~\ref{fig:method_tokenizer}.  

\subsection{Order-agnostic Item Tokenization} 
Meeting the two principles, SETRec leverages a CF and a semantic tokenizer to endow multi-dimensional information into a set of order-agnostic continuous tokens\footnote{We do not use discrete tokens in SETRec because discretization inevitably suffers from information loss~\cite{lazebnik2008supervised}, potentially leading to suboptimal results.} as illustrated in Figure~\ref{fig:method_tokenizer}(b). 

\noindent$\bullet\quad$\textbf{CF Tokenizer.}
As shown in Figure~\ref{fig:method_tokenizer}(b), we utilize a pre-trained conventional recommender model (\eg SASRec~\cite{lazebnik2008supervised}) with a linear projection layer to obtain item CF embedding $\bm{z}_{\text{CF}}\in\mathbb{R}^{d}$, where $d$ is the hidden dimension of LLMs. 
Incorporating CF embeddings encourages LLM-based recommenders to facilitate recommendations for users/items with rich interactions. 

\noindent$\bullet\quad$\textbf{Semantic Tokenizer.} 
To fully utilize rich item semantic information, SETRec introduces a semantic tokenizer to obtain a set of semantic embeddings. 
Specifically, 
given the item semantic information $c$ such as title and categories, we first extract the item semantic representations $\bm{s}$ with a pre-trained semantic extractor (\eg SentenceT5~\cite{ni2021sentence}). 

To obtain the semantic embeddings, a straightforward approach is to compress semantic representation $\bm{s}$ into a single latent semantic embedding. 
Nonetheless, compressing multi-dimensional semantic information (\eg ``brand'' and ``price'') might suffer from the embedding collapse issue~\cite{guoembedding,pan2024ads}, potentially undermining the rich semantic content that distinguishes between items. 
To prevent this issue, as depicted in Figure~\ref{fig:method_tokenizer}(b), we tokenize each item into $N$ order-agnostic semantic embeddings via an AE: 
\begin{equation}\small
\begin{aligned}
&\bm{z} = \text{Encoder}(\bm{s}), 
\end{aligned}
\end{equation}
where 
$\bm{z}=[\bm{z}_{S_{1}}, \bm{z}_{S_{2}},\dots,\bm{z}_{S_{N}}]\in\mathbb{R}^{Nd}$ denotes the concatenated semantic embeddings representing different latent semantic dimensions, and ${z}_{S_{n}}\in\mathbb{R}^{d}$ is the $n$-th semantic embedding. 
Notably, we utilize a unified AE instead of multiple independent AEs for two considerations: 
1) employing a single AE reduces the parameters with an approximate ratio of $\frac{1}{N}$, which is highly practical; 
2) alleviating the training instability that might be caused by multiple encoders' training~\cite{tang2023improving}. 
In addition, to encourage the semantic embeddings to preserve useful information as much as possible, 
a reconstruction loss is used to train the semantic tokenizer: 
\begin{equation}\small
    \mathcal{L}_{AE} = \|\bm{s}-\hat{\bm{s}}\|_2^{2}, 
\end{equation}
where 
$\hat{\bm{s}} = \text{Decoder}(\bm{z})$ is the reconstructed semantic representation. 

\noindent$\bullet\quad$\textbf{Token Corpus.} 
Based on the CF and the semantic tokenizer, we can obtain the set identifier for each item
$\tilde{i} = \{\bm{z}_{\text{CF}},\bm{z}_{S_{1}},\dots,\bm{z}_{S_{N}}\}$, consisting of a CF embedding and $N$ semantic embeddings. 
We then can collect tokens from all items and obtain the token corpus for each information dimension, \ie $\mathcal{Z}_\text{CF}, \mathcal{Z}_{S_1}, \dots, \mathcal{Z}_{S_N}$. 
The collected token corpus is used as the grounding head for effective item grounding (\cf Section~\ref{sec:query-based_decoding}). 

\subsection{Simultaneous Item Generation} 
To efficiently and effectively generate set identifiers, it is crucial for SETRec to 
1) guide LLMs to distinguish different dimensions and generate tokens aligning well with each dimension simultaneously (Section~\ref{sec:query-based_decoding}); 
2) ground the generated token set to existing items effectively (Section~\ref{sec:query-based_decoding}); 
3) eliminate the unnecessary dependencies introduced in user history (Section~\ref{sec:sparse_attention_mask}); 

\subsubsection{\textbf{Query-guided Generation}.}\label{sec:query-based_decoding}
As shown in Figure~\ref{fig:method_tokenizer}(a), to guide LLMs to generate tokens that align well with the information dimensions, 
we introduce a set of learnable query vectors $\bm{q}\in\mathbb{R}^{d}$, where $d$ is the latent dimension of the LLMs, to guide the LLMs to distinguish between information dimensions (\eg CF and semantic) for token generation. 
Formally, the generated token $\hat{\bm{z}}_k$ for each dimension $k\in\{\text{CF}, S_1, S_2, \dots, S_N\}$ is obtained via:
\begin{equation}\label{eqn:query_decoding}\small
\left\{
\begin{aligned}
    &\bm{x} = [\{\bm{z}_{\text{CF}}, \bm{z}_{S_1}, 
     \dots, \bm{z}_{S_N}\}^{1}, \dots, \{\bm{z}_{\text{CF}}, \bm{z}_{S_1}, \dots,\bm{z}_{S_N}\}^{L}], \\
    &\hat{\bm{z}}_{k} = \text{LLM\_Layers}(\bm{x}, \bm{q}_k),
\end{aligned}
\right.
\end{equation}
where $\bm{q}_k$ is the learnable query vector to guide LLM generation for the information dimension $k$. 
Based on Eq. (\ref{eqn:query_decoding}), we can collect the generated token for all dimensions and obtain the generated set identifier  $\hat{i}=\{\hat{\bm{z}}_{\text{CF}}, \hat{\bm{z}}_{S_1}, \dots, 
\hat{\bm{z}}_{S_N}
\}$. 

\vspace{2pt}
\textbf{\textit{Token Generation Optimization.}} 
To achieve accurate item recommendations, 
we encourage the generated token to align with the target token for every dimension: 
\begin{equation}\label{eqn:loss_gen}\small
    \mathcal{L}_{\text{Gen}} = - \frac{1}{|\mathcal{D}|} \sum_{\mathcal{D}} \sum_{k\in\mathcal{F}}\frac{\exp (sim(\hat{\bm{z}}_k,\bm{z}_{k}))}{\sum_{\bm{z}\in\mathcal{Z}_k}\exp(sim(\hat{\bm{z}}_k,\bm{z}))},
\end{equation}
where $\mathcal{F}=\{\text{CF}, S_1, \dots, S_N\}$, $sim(\cdot)$ is the similarity function (\eg inner product), and $\bm{z}_k$ is the target item token for the information dimension $k$. 
Intuitively, Eq. (\ref{eqn:loss_gen}) pushes the generated embedding closer to the target embedding and pulls away from other embeddings within the specific information dimension.  

\vspace{2pt}
\textit{\textbf{Token Generation Grounding.}} 
Based on generated tokens obtained via Eq. (\ref{eqn:query_decoding}), the next step is to ground them to the existing items. 
However, this can be challenging since the possible combinations of the tokens from different information dimensions are much larger than the existing item corpus, \ie $\prod_{k\in\mathcal{F}}|\mathcal{Z}_k|\gg |\mathcal{I}|$. 
To solve this issue, we introduce a token set grounding strategy, which leverages the token corpus as grounding heads to obtain the item score. 
Formally, we have 
\begin{equation}\label{eqn:single_logits}\small
\left\{
\begin{aligned}    &s_k=W_{k}\bm{\hat{z}}_k, \\
&s = (1-\beta) s_\text{CF} + \beta \sum\nolimits_{{k\in \mathcal{F}\setminus{\text{CF}}}} s_{k}, \\
\end{aligned}
\right.
\end{equation}
where $W_{k}\in\mathbb{R}^{|I|\times d}$ is adopted from the token corpus $\mathcal{Z}_k$. 
The final item scores are obtained via a linear combination of CF and semantic dimensions, where $\beta$ is a hyper-parameter to balance the strength between CF and semantic dimensions. 
It is highlighted that the grounding heads for semantic dimensions are extendable to new items, leading to strong generalization ability (\cf Section~\ref{sec:overall_performance}).

\begin{figure}[t]
\setlength{\abovecaptionskip}{0.02cm}
\setlength{\belowcaptionskip}{-0.3cm}
\centering
\includegraphics[scale=1.1]{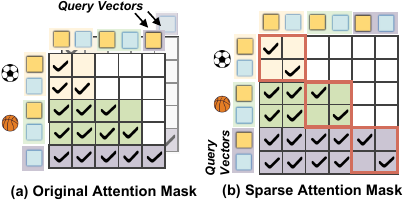}
\setlength{\fboxrule}{1pt}
\caption{Comparison between original attention and sparse attention ($N=1$). The sparse attention 1) eliminates the dependency over other tokens within the same item ({\color{myred}\fbox{\phantom{\rule{0.06cm}{0.06cm}}}}), and 2) boosts the efficiency with the flattened input, \ie query vectors are in the same sequence.}
\label{fig:sparse_attn}
\end{figure}

\vspace{2pt}
\subsubsection{\textbf{Sparse Attention Mask.}}\label{sec:sparse_attention_mask}
While simultaneous generation bypasses the sequential generation of item identifier, the flattened user's historical interactions are still sequentially encoded, inevitably introducing order information of tokens within each identifier (Figure~\ref{fig:sparse_attn}(a)). 
To combat this issue, we introduce a sparse attention mask as illustrated in Figure~\ref{fig:sparse_attn}(b). 
Specifically, for the user's historical interactions, tokens associated with an identifier are treated as independent from each other (\eg CF embedding cannot attend to semantic embeddings). 
However, these tokens can still attend to all tokens in previously interacted items (\eg a fully attended mask is applied to football when calculating self-attention for tokens in basketball).
Therefore, the sparse attention mask ensures the order agnosticism of the set identifier. 

\vspace{2pt}
\noindent$\bullet\quad$\textbf{Time Complexity Analysis.} 
Moreover, the sparse attention mask can improve the generation efficiency by reducing the duplicate computations of the shared prefix via original attention mask (Figure~\ref{fig:sparse_attn}(a)). 
With $M$ information dimensions and $L$ historically interacted items, the time complexity for batch generation with the original attention mask is $M^3L^2d$. 
Remarkably, based on the flattened input with our proposed sparse attention mask, the time complexity reduces to $M^2L^2d$. 





\subsection{Instantiation}
To instantiate SETRec on LLMs, we optimize the CF and semantic tokenizers, learnable query vectors, and LLMs by minimizing: 
\begin{equation}\label{eqn:overall_loss}\small
    \mathcal{L} = \mathcal{L}_{\text{Gen}} + \alpha \mathcal{L}_{\text{AE}},
\end{equation}
where $\alpha$ is a hyper-parameter to control the strength of the tokenizer training. 
During inference, SETRec first tokenizes all items into set identifiers and obtain token corpus $\mathcal{Z}$ for each information dimension. 
Then, to recommend item, SETRec transforms user history into identifier sequence and performs query-guided simultaneous generation with sparse attention mask via Eq. (\ref{eqn:query_decoding}) to generate tokens for all information dimensions.  
Finally, SETRec leverages token corpus as extendable grounding heads to ground the generated token set to the valid items via Eq. (\ref{eqn:single_logits}). 

\section{Experiment}\label{sec:experiment}
We carry out extensive experiments on four real-world datasets to answer the following research questions: 
\textbf{RQ1:} How does our proposed SETRec perform compared to different identifier baselines on different architectures of LLMs? 
\textbf{RQ2:} How do the different components of SETRec (\ie CF embeddings, semantic embeddings, query vectors, and sparse attention) affect the performance?
\textbf{RQ3:} How does SETRec perform when scaling up the model size and how does SETRec improve the overall performance? 
\textbf{RQ4:} How does SETRec perform with different number of semantic embeddings, tokenizer training strength, and semantic strength for inference? 


\subsection{Experimental Settings}
\subsubsection{\textbf{Datasets}}
We conduct experiments on four real-world datasets across various domains. 
From Amazon review datasets\footnote{\url{https://jmcauley.ucsd.edu/data/amazon/}.}, we adopt three widely used benchmarks 
1)\textbf{Toys}, 2) \textbf{Beauty}, and 3) \textbf{Sports}. 
The three Amazon datasets contain rich user interactions over a specific category of e-commerce products, where each item is associated with rich textual meta information such as title, description, category, and brand. 
In addition, we use a video games dataset 4) \textbf{Steam}\footnote{\url{https://github.com/kang205/SASRec}.} proposed in~\cite{kang2018self}, which contains substantial user interactions on video games with abundant textual semantic information. 
For all datasets, we follow previous work~\cite{wang2023causal} to sort user interactions chronologically according to the timestamps and divide them into training, validation, and testing sets with a ratio of 8:1:1. 
In addition, we divide the items into warm and cold items\footnote{We denote warm- and cold-start items as warm and cold items for brevity.}, where the items that appear in the training set are warm items, otherwise cold items.

\noindent$\bullet\quad$\textbf{Evaluation.} 
We adopt the widely used metrics Recall@$K$ and NDCG@$K$, where $K=5$ and $10$ to evaluate all methods. 
Additionally, 
we introduce three different settings that evaluate over 1) all items, 2) warm items only, and 3) cold items only, respectively.  

\begin{table*}[t]
\setlength{\abovecaptionskip}{0.05cm}
\setlength{\belowcaptionskip}{0.2cm}
\caption{Overall performance of baselines and SETRec instantiated on T5. The best results are in bold and the second-best results are underlined. $*$ implies the improvements over the second-best results are statistically significant ($p$-value < 0.01) under one-sample t-tests. ``Inf. Time'' denotes the inference time over all test users tested on a single NVIDIA RTX A5000 GPU.}
\setlength{\tabcolsep}{2mm}{
\resizebox{\textwidth}{!}{
\begin{tabular}{l|l|cccc|cccc|cccc|c}
\toprule
 &  & \multicolumn{4}{c|}{\textbf{All}} & \multicolumn{4}{c|}{\textbf{Warm}} & \multicolumn{4}{c|}{\textbf{Cold}} & \multicolumn{1}{l}{\textbf{Inf. Time (s)}} \\ \hline
\textbf{Dataset} & \textbf{Method} & \textbf{R@5} & \textbf{R@10} & \textbf{N@5} & \textbf{N@10} & \textbf{R@5} & \textbf{R@10} & \textbf{N@5} & \textbf{N@10} & \textbf{R@5} & \textbf{R@10} & \textbf{N@5} & \textbf{N@10} & \textbf{All Users} \\ \midrule
\multirow{9}{*}{\textbf{Toys}} & \textbf{DreamRec} & 0.0020 & 0.0027 & 0.0015 & 0.0018 & 0.0027 & 0.0039 & 0.0020 & 0.0024 & 0.0066 & 0.0168 & 0.0045 & 0.0082 & 912 \\
 & \textbf{E4SRec} & 0.0061 & 0.0098 & 0.0051 & 0.0064 & 0.0081 & 0.0128 & 0.0065 & 0.0082 & 0.0065 & 0.0122 & 0.0056 & 0.0078 & \textbf{55} \\ \cmidrule{2-15}
 & \textbf{BIGRec} & 0.0008 & 0.0013 & 0.0007 & 0.0009 & 0.0014 & 0.0019 & 0.0011 & 0.0013 & 0.0278 & 0.0360 & 0.0196 & 0.0223 & 2,079 \\
 & \textbf{IDGenRec} & 0.0063 & 0.0110 & 0.0052 & 0.0069 & 0.0109 & {\ul 0.0161} & 0.0081 & {0.0102} & {\ul 0.0318} & {\ul 0.0589} & {\ul 0.0236} & {\ul 0.0335} & 658 \\
 & \textbf{CID} & 0.0044 & 0.0082 & 0.0040 & 0.0053 & 0.0065 & 0.0128 & 0.0049 & 0.0071 & 0.0059 & 0.0111 & 0.0047 & 0.0066 & 810 \\
 & \textbf{SemID} & 0.0071 & 0.0108 & 0.0061 & 0.0074 & 0.0086 & 0.0153 & 0.0075 & 0.0100 & 0.0307 & 0.0507 & 0.0220 & 0.0292 & 1,215 \\
 & \textbf{TIGER} & 0.0064 & 0.0106 & 0.0060 & 0.0076 & 0.0091 & 0.0147 & 0.0080 & {\ul 0.0102} & 0.0315 & 0.0555 & 0.0228 & 0.0314 & 448 \\
 & \textbf{LETTER} & {\ul 0.0081} & {\ul 0.0117} & {\ul 0.0064} & {\ul 0.0077} & {\ul 0.0109} & 0.0155 & {\ul 0.0083} & 0.0101 & 0.0183 & 0.0395 & 0.0115 & 0.0190 & 448 \\  \cmidrule{2-15}
 & \cellcolor{gray!16}\textbf{SETRec} & \cellcolor{gray!16}\textbf{0.0110*} & \cellcolor{gray!16}\textbf{0.0189*} & \cellcolor{gray!16}\textbf{0.0089*} & \cellcolor{gray!16}\textbf{0.0118*} & \cellcolor{gray!16}\textbf{0.0139*} & \cellcolor{gray!16}\textbf{0.0236*} & \cellcolor{gray!16}\textbf{0.0112*} & \cellcolor{gray!16}\textbf{0.0147*} & \cellcolor{gray!16}\textbf{0.0443*} & \cellcolor{gray!16}\textbf{0.0812*} & \cellcolor{gray!16}\textbf{0.0310*} & \cellcolor{gray!16}\textbf{0.0445*} & \cellcolor{gray!16}{\ul 60} \\ \midrule\midrule
\multirow{9}{*}{\textbf{Beauty}} & \textbf{DreamRec} & 0.0012 & 0.0025 & 0.0013 & 0.0017 & 0.0016 & 0.0028 & 0.0016 & 0.0019 & 0.0078 & 0.0161 & 0.0065 & 0.0094 & 1,102 \\
 & \textbf{E4SRec} & 0.0061 & 0.0092 & 0.0052 & 0.0063 & 0.0080 & 0.0121 & 0.0067 & 0.0082 & 0.0072 & 0.0118 & 0.0065 & 0.0077 & \textbf{120} \\ \cmidrule{2-15}
 & \textbf{BIGRec} & 0.0054 & 0.0064 & 0.0051 & 0.0054 & 0.0008 & 0.0009 & 0.0006 & 0.0008 & 0.0106 & 0.0251 & 0.0095 & 0.0151 & 4,544 \\
 & \textbf{IDGenRec} & {\ul 0.0080} & 0.0115 & {\ul 0.0066} & {0.0078} & {\ul 0.0106} & 0.0165 & 0.0078 & 0.0099 & 0.0187 & 0.0350 & 0.0186 & 0.0224 & 840 \\
 & \textbf{CID} & 0.0071 & 0.0125 & 0.0060 & {\ul 0.0080} & 0.0098 & {0.0166} & 0.0077 & 0.0101 & 0.0087 & 0.0183 & 0.0071 & 0.0104 & 815 \\
 & \textbf{SemID} & 0.0071 & {\ul 0.0131} & 0.0056 & {0.0078} & 0.0098 & {\ul 0.0174} & 0.0074 & {\ul 0.0103} & {\ul 0.0260} & {\ul 0.0465} & 0.0178 & 0.0255 & 1,310 \\
 & \textbf{TIGER} & 0.0063 & 0.0098 & 0.0050 & 0.0062 & 0.0086 & 0.0131 & 0.0065 & 0.0082 & 0.0190 & 0.0325 & 0.0130 & 0.0178 & 430 \\ 
 & \textbf{LETTER} & 0.0071 & 0.0103 & 0.0061 & 0.0070 & 0.0094 & 0.0135 & {\ul 0.0079} & 0.0091 & 0.0251 & 0.0410 & {\ul 0.0241} & {\ul 0.0285} & 430 \\ \cmidrule{2-15}
 & \cellcolor{gray!16}\textbf{SETRec} & \cellcolor{gray!16}\textbf{0.0106*} & \cellcolor{gray!16}\textbf{0.0161*} & \cellcolor{gray!16}\textbf{0.0083*} & \cellcolor{gray!16}\textbf{0.0103*} & \cellcolor{gray!16}\textbf{0.0139*} & \cellcolor{gray!16}\textbf{0.0212*} & \cellcolor{gray!16}\textbf{0.0108*} & \cellcolor{gray!16}\textbf{0.0134*} & \cellcolor{gray!16}\textbf{0.0384*} & \cellcolor{gray!16}\textbf{0.0761*} & \cellcolor{gray!16}\textbf{0.0280*} & \cellcolor{gray!16}\textbf{0.0413*} & \cellcolor{gray!16}{\ul 126} \\ \midrule\midrule
\multirow{9}{*}{\textbf{Sports}} & \textbf{DreamRec} & 0.0027 & 0.0044 & 0.0025 & 0.0031 & 0.0032 & 0.0052 & 0.0028 & 0.0035 & 0.0045 & 0.0108 & 0.0026 & 0.0049 & 2,100 \\ 
 & \textbf{E4SRec} & 0.0079 & 0.0131 & 0.0075 & 0.0094 & 0.0092 & 0.0154 & 0.0085 & 0.0107 & 0.0031 & 0.0093 & 0.0019 & 0.0039 & \textbf{117} \\ \cmidrule{2-15}
 & \textbf{BIGRec} & 0.0033 & 0.0042 & 0.0030 & 0.0033 & 0.0001 & 0.0002 & 0.0001 & 0.0001 & 0.0059 & 0.0104 & 0.0043 & 0.0061 & 7,822 \\
 & \textbf{IDGenRec} & 0.0087 & 0.0127 & 0.0079 & 0.0092 & 0.0101 & 0.0149 & 0.0091 & 0.0107 & 0.0181 & 0.0302 & 0.0134 & 0.0179 & 1,724 \\
 & \textbf{CID} & 0.0077 & 0.0131 & 0.0073 & 0.0092 & 0.0074 & 0.0119 & 0.0045 & 0.0061 & 0.0082 & 0.0149 & 0.0075 & 0.0099 & 2,135 \\
 & \textbf{SemID} & {\ul 0.0094} & {\ul 0.0167} & {\ul 0.0088} & {\ul 0.0114} & {\ul 0.0119} & {\ul 0.0201} & {\ul 0.0104} & {\ul 0.0135} & {\ul 0.0254} & {\ul 0.0495} & {\ul 0.0175} & {\ul 0.0256} & 2,367 \\
 & \textbf{TIGER} & 0.0085 & 0.0129 & 0.0080 & 0.0095 & 0.0100 & 0.0151 & 0.0091 & 0.0109 & 0.0190 & 0.0310 & 0.0120 & 0.0159 & 481 \\
 & \textbf{LETTER} & 0.0077 & 0.0131 & 0.0073 & 0.0092 & 0.0074 & 0.0119 & 0.0045 & 0.0061 & 0.0082 & 0.0149 & 0.0075 & 0.0099 & 481 \\ \cmidrule{2-15}
 & \cellcolor{gray!16}\textbf{SETRec} & \cellcolor{gray!16}\textbf{0.0114*} & \cellcolor{gray!16}\textbf{0.0185*} & \cellcolor{gray!16}\textbf{0.0101*} & \cellcolor{gray!16}\textbf{0.0126*} & \cellcolor{gray!16}\textbf{0.0134*} & \cellcolor{gray!16}\textbf{0.0216*} & \cellcolor{gray!16}\textbf{0.0115*} & \cellcolor{gray!16}\textbf{0.0144*} & \cellcolor{gray!16}\textbf{0.0341*} & \cellcolor{gray!16}\textbf{0.0595*} & \cellcolor{gray!16}\textbf{0.0233*} & \cellcolor{gray!16}\textbf{0.0323*} & \cellcolor{gray!16}{\ul 136} \\ \midrule\midrule
\multirow{9}{*}{\textbf{Steam}} & \textbf{DreamRec} & 0.0029 & 0.0057 & 0.0037 & 0.0046 & 0.0042 & 0.0080 & 0.0045 & 0.0059 & 0.0017 & 0.0029 & 0.0013 & 0.0018 & 4,620 \\
 & \textbf{E4SRec} & 0.0194 & 0.0351 & 0.0220 & 0.0270 & 0.0312 & 0.0558 & 0.0283 & 0.0370 & 0.0006 & 0.0010 & 0.0006 & 0.0007 & \textbf{328} \\ \cmidrule{2-15}
 & \textbf{BIGRec} & 0.0030 & 0.0049 & 0.0046 & 0.0049 & 0.0048 & 0.0053 & 0.0061 & 0.0053 & 0.0099 & 0.0107 & {\ul 0.0129} & 0.0127 & 5,167 \\
 & \textbf{IDGenRec} & 0.0199 & 0.0307 & 0.0241 & 0.0265 & 0.0309 & 0.0479 & 0.0311 & 0.0363 & 0.0047 & 0.0151 & 0.0039 & 0.0078 & 2,846 \\
 & \textbf{CID} & 0.0200 & {\ul 0.0360} & {\ul 0.0249} & {\ul 0.0295} & 0.0314 & {\ul 0.0566} & {\ul 0.0315} & {\ul 0.0400} & 0.0008 & 0.0021 & 0.0006 & 0.0011 & 3,194 \\
 & \textbf{SemID} & 0.0155 & 0.0278 & 0.0192 & 0.0229 & 0.0248 & 0.0443 & 0.0246 & 0.0313 & 0.0017 & 0.0027 & 0.0015 & 0.0018 & 3,605 \\
 & \textbf{TIGER} & {\ul 0.0202} & 0.0348 & 0.0244 & 0.0287 & {\ul 0.0320} & 0.0552 & 0.0314 & 0.0393 & 0.0060 & {0.0152} & 0.0044 & 0.0078 & 1,747 \\
 & \textbf{LETTER} & 0.0164 & 0.0312 & 0.0195 & 0.0244 & 0.0268 & 0.0500 & 0.0253 & 0.0336 & {\ul 0.0115} & {\ul 0.0317} & {0.0077} & {\ul 0.0157} & 1,747 \\ \cmidrule{2-15}
 & \cellcolor{gray!16}\textbf{SETRec} & \cellcolor{gray!16}\textbf{0.0216*} & \cellcolor{gray!16}\textbf{0.0383*} & \cellcolor{gray!16}\textbf{0.0254*} & \cellcolor{gray!16}\textbf{0.0308*} & \cellcolor{gray!16}\textbf{0.0339*} & \cellcolor{gray!16}\textbf{0.0591*} & \cellcolor{gray!16}\textbf{0.0326*} & \cellcolor{gray!16}\textbf{0.0414*} & \cellcolor{gray!16}\textbf{0.0313*} & \cellcolor{gray!16}\textbf{0.0572*} & \cellcolor{gray!16}\textbf{0.0248*} & \cellcolor{gray!16}\textbf{0.0342*} & \cellcolor{gray!16}{\ul 347} \\ \hline
\end{tabular}
}}
\label{tab:overall_performance}
\end{table*}

\subsubsection{\textbf{Baselines}}
We compare SETRec with competitive baselines, including single-token identifiers (DreamRec, E4SRec) and token-sequence identifiers (BIGRec, IDGenRec, CID, SemID, TIGER, LETTER). 
1) \textbf{DreamRec}~\cite{yang2024generate} is a closely related method that leverages ID embedding to represent each item and adopts a diffusion model to refine the generated ID embedding from LLMs.  
2) \textbf{E4SRec}~\cite{li2023e4srec} utilizes a pre-trained CF model to obtain ID embedding, and uses a linear projection layer to obtain the item scores efficiently. 
3) \textbf{BIGRec}~\cite{bao2023bi} adopts item titles as identifiers, where the tokens are from human vocabulary. 
4) \textbf{IDGenRec}~\cite{tan2024idgenrec} is a learnable ID generator, which aims to generate concise but informative tags from human vocabulary to represent each item. 
5) \textbf{CID}~\cite{hua2023index} leverages hierarchical clustering to obtain token sequence, which utilizes item co-occurrence matrix to obtain identifiers to ensure items with similar interactions share similar tokens. 
6) \textbf{SemID}~\cite{hua2023index} also represents items with external token sequence, which is obtained based on the hierarchical item category. 
7) \textbf{TIGER}~\cite{rajput2023recommender} leverages RQ-VAE with codebooks to quantize item semantic information into token sequence with external tokens. The identifier sequentially contains coarse-grained to fine-grained information. 
8) \textbf{LETTER}~\cite{wang2024learnable} is one of the SOTA item tokenization methods, which incorporates both semantic and CF information into the training of RQ-VAE, achieving identifiers with multi-dimensional information and improved diversity. 

\subsubsection{\textbf{Implementation Details}} 
We instantiate all methods on two LLMs with different architectures, \ie T5-small~\cite{raffel2020exploring} (encoder-decoder) and Qwen2.5~\cite{yang2024qwen2} (decoder-only). 
Specifically, we adopt Qwen\footnote{We denote T5-small and Qwen2.5 as T5 and Qwen for brevity.} with different sizes, including 1.5B, 3B, and 7B, for a comprehensive evaluation. 
To ensure a fair comparison, we set the hidden layer dimensions at 512, 256, and 128 with ReLU activation for methods that adopt AE in tokenizer training, including TIGER, LETTER, and our proposed SETRec. 
For LLM training, 
we adopt the same prompt for all methods as ``What would the user be likely to purchase next after buying items {history}?;'' for a fair comparison. 
We fully fine-tune the T5 model and perform parameter-efficient fine-tuning technique LoRA~\cite{hu2021lora} for Qwen. 
All experiments are conducted on four NVIDIA RTX A5000 GPUs. 
For SETRec, 
we select $N$, $\alpha$, and $\beta$ from $\{1,2,3,4,5,6\}$, $\{0.1,0.3,0.5,0.7,0.9\}$, and $\{0, 0.1, 0.2, 0.3, 0.4, 0.5, 0.6, 0.7, 0.8, 0.9,1.0\}$, respectively.

\begin{table*}[t]
\setlength{\abovecaptionskip}{0.05cm}
\setlength{\belowcaptionskip}{0.2cm}
\caption{Overall performance on Qwen-1.5B over Toys and Beauty. The best results are in bold and the second-best results are underlined. ``Inf. Time'' denotes the inference time over all test users tested on a single NVIDIA RTX A5000 GPU.}
\setlength{\tabcolsep}{2mm}{
\resizebox{\textwidth}{!}{
\begin{tabular}{l|l|cccc|cccc|cccc|c}
\toprule
 &  & \multicolumn{4}{c}{\textbf{All}} & \multicolumn{4}{c}{\textbf{Warm}} & \multicolumn{4}{c}{\textbf{Cold}} & \textbf{Inf. Time(s)} \\ \hline
\textbf{Dataset} & \textbf{Method} & \textbf{R@5} & \textbf{R@10} & \textbf{N@5} & \textbf{N@10} & \textbf{R@5} & \textbf{R@10} & \textbf{N@5} & \textbf{N@10} & \textbf{R@5} & \textbf{R@10} & \textbf{N@5} & \textbf{N@10} & \textbf{All Users} \\ \midrule
\multirow{9}{*}{\textbf{Toys}} & \textbf{DreamRec} & 0.0006 & 0.0013 & 0.0005 & 0.0008 & 0.0008 & 0.0019 & 0.0007 & 0.0012 & 0.0076 & 0.0137 & 0.0052 & 0.0074 & 1,093 \\
 & \textbf{E4SRec} & 0.0065 & 0.0108 & {\ul 0.0056} & 0.0072 & 0.0089 & 0.0144 & {\ul 0.0075} & {\ul 0.0096} & 0.0084 & 0.0235 & 0.0055 & 0.0111 & \textbf{905} \\ \cmidrule{2-15} 
 & \textbf{BIGRec} & 0.0009 & 0.0016 & 0.0009 & 0.0012 & 0.0011 & 0.0013 & 0.0010 & 0.0011 & 0.0194 & 0.0311 & 0.0147 & 0.0191 & 43,304 \\
 & \textbf{IDGenRec} & 0.0030 & 0.0053 & 0.0022 & 0.0031 & 0.0043 & 0.0086 & 0.0032 & 0.0048 & 0.0189 & 0.0364 & 0.0161 & 0.0224 & 30,720 \\
 & \textbf{CID} & 0.0027 & 0.0047 & 0.0025 & 0.0033 & 0.0055 & 0.0084 & 0.0044 & 0.0056 & 0.0055 & 0.0156 & 0.0044 & 0.0081 & {27,248} \\
 & \textbf{SemID} & 0.0024 & 0.0042 & 0.0018 & 0.0024 & 0.0034 & 0.0055 & 0.0026 & 0.0034 & 0.0140 & 0.0275 & 0.0095 & 0.0143 & 32,288 \\
 & \textbf{TIGER} & {\ul 0.0068} & {\ul 0.0117} & 0.0054 & {\ul 0.0072} & {\ul 0.0094} & {\ul 0.0159} & 0.0070 & 0.0095 & {\ul 0.0384} & {\ul 0.0715} & {\ul 0.0291} & {\ul 0.0408} & {13,800} \\
 & \textbf{LETTER} & 0.0057 & 0.0093 & 0.0050 & 0.0064 & 0.0080 & 0.0126 & 0.0066 & 0.0085 & 0.0217 & 0.0416 & 0.0170 & 0.0239 & 13,800 \\ \cmidrule{2-15} 
 & \cellcolor[HTML]{ECF4FF}\textbf{SETRec} & \cellcolor[HTML]{ECF4FF}\textbf{0.0116*} & \cellcolor[HTML]{ECF4FF}\textbf{0.0188*} & \cellcolor[HTML]{ECF4FF}\textbf{0.0095*} & \cellcolor[HTML]{ECF4FF}\textbf{0.0120*} & \cellcolor[HTML]{ECF4FF}\textbf{0.0144*} & \cellcolor[HTML]{ECF4FF}\textbf{0.0236*} & \cellcolor[HTML]{ECF4FF}\textbf{0.0118*} & \cellcolor[HTML]{ECF4FF}\textbf{0.0151*} & \cellcolor[HTML]{ECF4FF}\textbf{0.0531*} & \cellcolor[HTML]{ECF4FF}\textbf{0.0883*} & \cellcolor[HTML]{ECF4FF}\textbf{0.0382*} & \cellcolor[HTML]{ECF4FF}\textbf{0.0507*} & \cellcolor[HTML]{ECF4FF}{\ul 926} \\ \midrule\midrule
\multirow{9}{*}{\textbf{Beauty}} & \textbf{DreamRec} & 0.0007 & 0.0009 & 0.0005 & 0.0005 & 0.0010 & 0.0011 & 0.0007 & 0.0007 & 0.0090 & 0.0167 & 0.0075 & 0.0103 & 1,326 \\
 & \textbf{E4SRec} & {\ul 0.0067} & {\ul 0.0109} & {\ul 0.0056} & {\ul 0.0072} & {\ul 0.0088} & {\ul 0.0146} & {\ul 0.0072} & {\ul 0.0094} & 0.0017 & 0.0071 & 0.0010 & 0.0029 & \textbf{910} \\ \cmidrule{2-15} 
 & \textbf{BIGRec} & 0.0006 & 0.0010 & 0.0006 & 0.0007 & 0.0010 & 0.0010 & 0.0008 & 0.0008 & 0.0141 & 0.0246 & 0.0094 & 0.0135 & 29,500 \\
 & \textbf{IDGenRec}  & 0.0042 & 0.0078 & 0.0030 & 0.0043 & 0.0045 & 0.0104 & 0.0033 & 0.0054 & {\ul 0.0254} & {\ul 0.0471} & {\ul 0.0207} & {\ul 0.0292} & 35,040 \\
 & \textbf{CID} & 0.0046 & 0.0077 & 0.0040 & 0.0052 & 0.0059 & 0.0107 & 0.0051 & 0.0068 & 0.0075 & 0.0155 & 0.0071 & 0.0096 & {27,792} \\
 & \textbf{SemID} & 0.0030 & 0.0045 & 0.0027 & 0.0033 & 0.0050 & 0.0076 & 0.0042 & 0.0052 & 0.0159 & 0.0227 & 0.0116 & 0.0159 & 45,160 \\
 & \textbf{TIGER} & 0.0041 & 0.0065 & 0.0032 & 0.0041 & 0.0054 & 0.0085 & 0.0042 & 0.0054 & 0.0083 & 0.0167 & 0.0064 & 0.0091 & {12,600} \\
 & \textbf{LETTER} & 0.0040 & 0.0069 & 0.0031 & 0.0042 & 0.0051 & 0.0088 & 0.0039 & 0.0054 & 0.0043 & 0.0129 & 0.0043 & 0.0071 & 12,600 \\ \cmidrule{2-15} 
 & \cellcolor[HTML]{ECF4FF}\textbf{SETRec} & \cellcolor[HTML]{ECF4FF}\textbf{0.0104*} & \cellcolor[HTML]{ECF4FF}\textbf{0.0167*} & \cellcolor[HTML]{ECF4FF}\textbf{0.0085*} & \cellcolor[HTML]{ECF4FF}\textbf{0.0108*} & \cellcolor[HTML]{ECF4FF}\textbf{0.0140*} & \cellcolor[HTML]{ECF4FF}\textbf{0.0221*} & \cellcolor[HTML]{ECF4FF}\textbf{0.0109*} & \cellcolor[HTML]{ECF4FF}\textbf{0.0141*} & \cellcolor[HTML]{ECF4FF}\textbf{0.0477*} & \cellcolor[HTML]{ECF4FF}\textbf{0.0748*} & \cellcolor[HTML]{ECF4FF}\textbf{0.0370*} & \cellcolor[HTML]{ECF4FF}\textbf{0.0464*} & \cellcolor[HTML]{ECF4FF}{\ul 1,050} \\ \bottomrule
\end{tabular}
}}
\label{tab:Overall_performance_on_Qwen}
\end{table*}

\subsection{Overall Performance (RQ1)}\label{sec:overall_performance}

\subsubsection{\textbf{Performance on T5.}} 
The performance comparison between baselines and SETRec instantiated on T5 are shown in Table~\ref{tab:overall_performance}, from which we have the following observations: 
\begin{itemize}[leftmargin=*]
    \item Token-sequence identifier (BIGRec, IDGenRec, CID, SemID, TIGER, LETTER) generally performs better than single-token identifier under ``all'', ``warm'', and ``cold'' settings. This is reasonable because token-sequence identifier represent each item with multiple tokens, which explicitly encode rich item information into different dimensions.   
    \item Among the token-sequence identifiers, methods with external tokens (CID, SemID, TIGER, LETTER) generally outperform those relying on human vocabulary (\eg BIGRec) under ``all'' and ``warm'' settings. 
    This is attributed to their hierarchically structured identifier, where the initial tokens represent coarse-grained semantics while subsequent tokens contain fine-grained semantics. 
    This aligns better with the autoregressive generation process, potentially alleviating the local optima issue~\cite{wang2024learnable}. 
    \item When recommending cold items\footnote{The higher values on cold performance are due to the limited number of cold items.}, methods that merely utilize CF information (DreamRec, E4SRec, and CID) fail to give satisfying results. 
    This is not surprising since CF information depends heavily on substantial interactions for training, thereby struggling with cold items. 
    In contrast, methods that integrate semantics into identifiers (BIGRec, IDGenRec, SemID, TIGER, and LETTER) generalize better on cold-start scenarios (superior performance under ``cold'' setting). 
    Specifically, BIGRec and IDGenRec tend to have competitive performance. 
    This is reasonable because they utilize readable human vocabulary to represent each item, which better leverages rich world knowledge encoded in LLMs. 

    \item SETRec significantly outperform all baselines under ``all'', ``warm'', and ``cold'' settings across all four datasets. 
    The superior performance is attributed to 
    1) the incorporation of both CF and semantic information into a set of tokens, which ensures accurate warm item recommendation and strong generalization on cold items; 
    2) order agnosticism of identifier, which removes the possibly inaccurate dependencies across different tokens associated with an identifier. 
    
    \item From the perspective of efficiency, SETRec significantly reduces the inference time costs compared to the token-sequence identifiers. 
    SETRec achieves an average 15$\times$, 11$\times$, 18$\times$, and 8$\times$ speedup on Toys, Beauty, Sports, and Steam, respectively, compared to token-sequence identifiers. 
    The high efficiency is attributed to the simultaneous generation, which generates multiple tokens at a single LLM call, unlocking the real-world deployment of LLM-based generative recommendation. 
\end{itemize}



\subsubsection{\textbf{Performance on Qwen-1.5B}}
To evaluate SETRec on decoder-only LLMs, we instantiate SETRec and all baselines on Qwen-1.5B. We present the results on Toys and Beauty\footnote{We omit the results with similar observations on other datasets to save space.} in Table~\ref{tab:Overall_performance_on_Qwen}, from which we summarize several key different observations from performance on T5 as follows: 

\begin{itemize}[leftmargin=*]
    \item Token-sequence identifiers show limited competitiveness compared to the counterparts on T5. 
    A possible reason is that Qwen-1.5B probably contains richer knowledge within its parameters, which amplifies the knowledge gap between the pre-training and recommendation tasks,  thereby hindering its adaptation to recommendation tasks with limited interaction data.  
    Conversely, E4SRec yields competitive performance in most cases. 
    This makes sense because E4SRec removes the original vocabulary head and replaces it with an item projection head, thus facilitating effective adaption to the recommendation tasks. 
    \item BIGRec and IDGenRec outperform their T5 counterparts on cold items on Beauty. Because they represent items with human vocabulary, which can leverage the rich world knowledge within Qwen-1.5B for better generalization. 
    On the contrary, identifiers with external tokens have inferior cold performance compared to their T5 counterparts. 
    This is also reasonable since it requires extensive interaction data to train external tokens. Otherwise, it is difficult for it to generalize to cold items accurately due to the low generation probability of these external tokens. 
    \item SETRec constantly outperforms baselines, which is consistent with the observations on T5. 
    Notably, SETRec instantiated on Qwen-1.5B steadily surpasses SETRec on T5, especially under the ``cold'' setting. 
    This validates the strong generalization ability of SETRec on different architectures of LLMs. 
    Moreover, as the LLM size increases, the efficiency improvements over the token-sequence identifiers are more significant, resulting in an average of 20$\times$ speedup across the two datasets. 
    
\end{itemize}

\subsection{In-depth Analysis}

\subsubsection{\textbf{Ablation Study (RQ2)}} 
\begin{figure}[t]
\setlength{\abovecaptionskip}{0.02cm}
\setlength{\belowcaptionskip}{-0.3cm}
\centering
\includegraphics[scale=1.2]{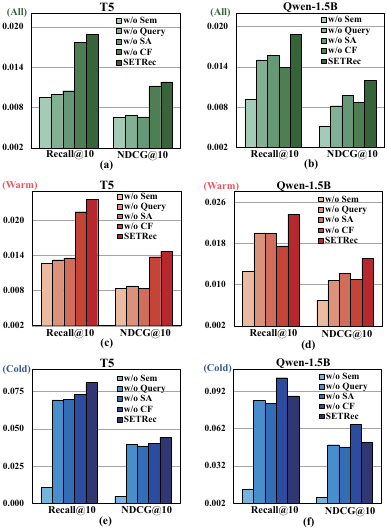}
\caption{Ablation study on Toys.}
\label{fig:ablation}
\end{figure}

To study the effectiveness of each component of SETRec, we separately remove semantic tokens (``w/o Sem''), 
CF token (``w/o CF'').  
In addition, we replace learnable query vectors with random frozen vectors (``w/o Query'') and 
use the original attention mask (``w/o SA''), to evaluate the effect of query vectors and the sparse attention mask, respectively. 
The results of different ablation variants on T5 and Qwen-1.5B on Toys are presented in Figure~\ref{fig:ablation} and we omit the results on other datasets with similar observations to save space. 

From the figures, we can find similar observations on T5 and Qwen that 
1) removing each component causes performance drops under ``all'', ``warm'', and ``cold'' settings, which validates the effectiveness of each component of SETRec. 
2) Discarding semantic tokens drastically degrades the recommendation accuracy under ``cold'' settings. 
This demonstrates the necessity of incorporating semantics into identifiers. 
Interestingly, 
3) removing semantic tokens leads to worse performance compared to removing CF token. 
The possible reason for this is the utilization of multiple semantic tokens to represent each item, which highlights the significance of leveraging multi-dimensional semantic information. 
This observation is also consistent with the results in~\cite{lin2024bridging}. 
Nonetheless, 
4) while removing CF tokens for T5 leads to inferior performance on cold items, using CF tokens for Qwen might negatively impact on cold items. 
A possible reason is that the larger-size Qwen is better at understanding semantics due to its stronger knowledge base encoded in the parameters, making the contribution of CF less significant.


\subsubsection{\textbf{Item Group Analysis (RQ3)}}
\begin{figure}[t]
\vspace{-0.2cm}
\setlength{\abovecaptionskip}{-0.15cm}
\setlength{\belowcaptionskip}{-0cm}
  \centering 
  \subfigure{
    \includegraphics[height=1.65in]{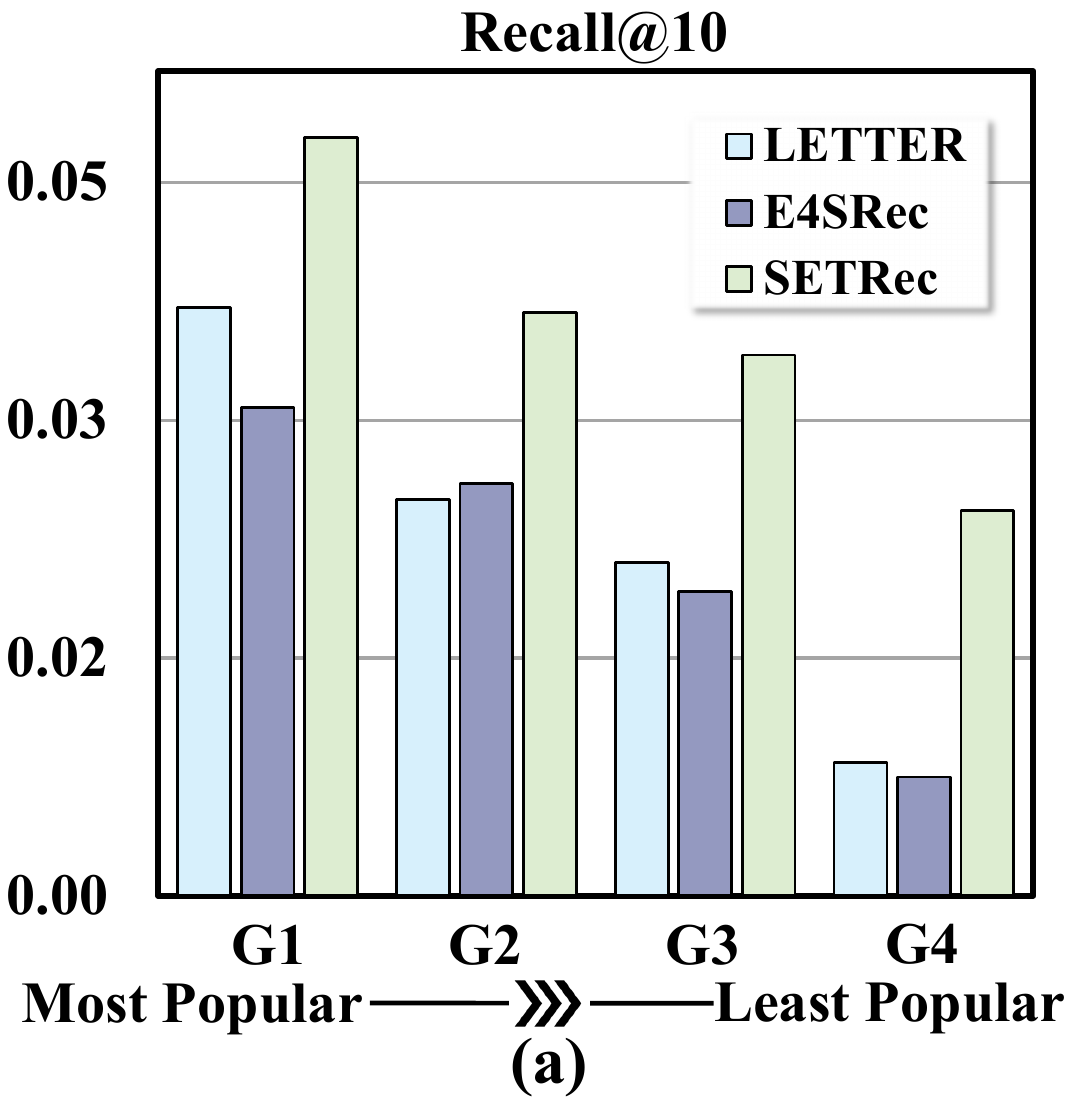}} 
  \subfigure{
    \includegraphics[height=1.65in]{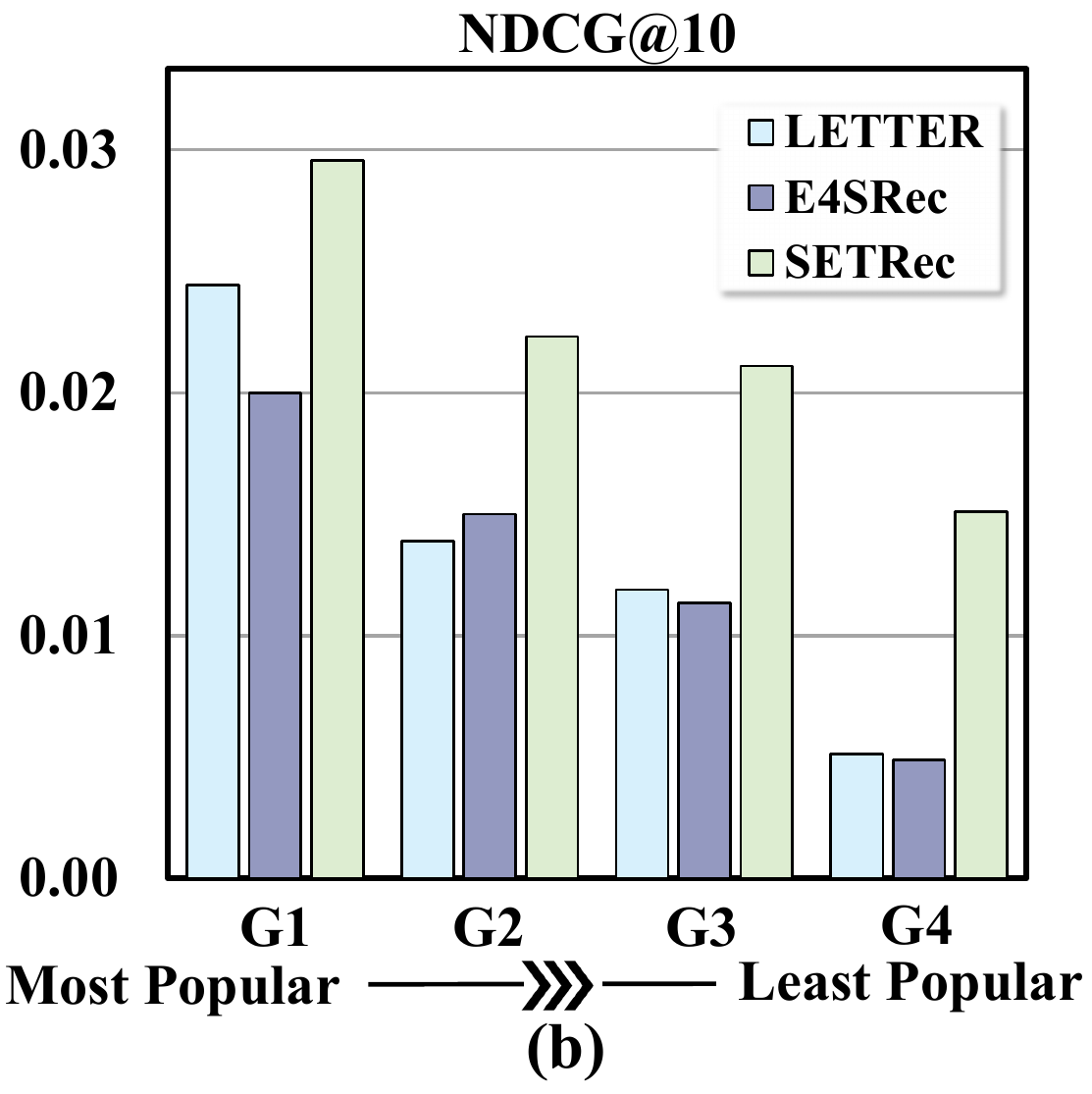}} 
\caption{Performance of SETRec, LETTER, and E4SRec (T5) on item groups with different popularity on Toys.}
  \label{fig:group_analysis}
\end{figure}

To understand how SETRec improves performance, we evaluate it over items with different popularity. 
The performance comparison between SETRec and two competitive baselines from token-sequence identifiers (LETTER) and single-token identifiers (E4SRec) are reported in Figure~\ref{fig:group_analysis}. 
We can observe that 
1) the performance gradually drops from G1 to G4. 
This makes sense since the less popular items have fewer interactions for LLMs to learn, thus leading to worse generation probabilities. 
Besides, 
2) E4SRec outperforms LETTER on most popular items (G1) but usually yields inferior performance on unpopular items (G2-G4). 
This is due to that E4SRec only uses CF information, which relies on substantial interactions and therefore struggle on unpopular items. 
In contrast, LETTER additionally incorporates semantics into identifiers, thus achieving better generalization on sparse items. 
3) SETRec consistently excels both E4SRec and LETTER over all groups. 
Notably, the improvements over sparse items are more significant, which partially explains the superiority of SETRec regarding overall performance.

\subsubsection{\textbf{Scalability on Model Parameters (RQ3)}}
To investigate whether SETRec can bring continuous performance when expanding the model parameters, we test SETRec on Qwen with different model sizes (1.5B, 3B, and 7B). 
Performance comparisons between SETRec, E4SRec, and LETTER on Toys are shown in Table~\ref{tab:scaling_performance}. 
From the results, we can find that 
1) SETRec clearly shows continued improvements over cold-start items when the model size scales from 1.5B to 7B, demonstrating promising scalability on cold items. 
We attribute this to the continued improvements of better semantic understanding by expanding the model parameters. 
Nonetheless, 
2) the performance on the warm items fails to continuously improve, indicating a relatively limited scalability over warm items. 
This shows that the larger models do not necessarily lead to better CF information understanding, which can also be indicated by the limited improvements of E4SRec under ``warm'' setting. 
Besides, 
3) LETTER shows weak scalability over the three settings. 
This is mainly due to the utilization of external tokens, which do not necessarily align with the pre-trained knowledge in LLMs, thus showing limited improvements by expanding the model parameters.

\begin{table}[t]
\setlength{\abovecaptionskip}{0.05cm}
\setlength{\belowcaptionskip}{0.2cm}
\caption{Performance comparison between SETRec and competitive baselines with different LLM sizes on Qwen. }
\setlength{\tabcolsep}{2.2mm}{
\resizebox{0.46\textwidth}{!}{
\begin{tabular}{clcccccc}
\toprule
\multicolumn{1}{l|}{} & \multicolumn{1}{l|}{} & \multicolumn{2}{c}{\textbf{All}} & \multicolumn{2}{c}{\textbf{Warm}} & \multicolumn{2}{c}{\textbf{Cold}} \\
\multicolumn{1}{l|}{} & \multicolumn{1}{l|}{} & \textbf{R@10} & \textbf{N@10} & \textbf{R@10} & \textbf{N@10} & \textbf{R@10} & \textbf{N@10} \\ \midrule\midrule
\multicolumn{1}{c|}{\multirow{3}{*}{\textbf{1.5B}}} & \multicolumn{1}{l|}{\textbf{LETTER}} & 0.0093 & 0.0064 & 0.0126 & 0.0085 & 0.0416 & 0.0239 \\
\multicolumn{1}{c|}{} & \multicolumn{1}{l|}{\textbf{E4SRec}} & 0.0108 & 0.0072 & 0.0144 & 0.0096 & 0.0235 & 0.0111 \\
\multicolumn{1}{c|}{} & \multicolumn{1}{l|}{\cellcolor{gray!16}\textbf{SETRec}} & \cellcolor{gray!16}\textbf{0.0188} & \cellcolor{gray!16}\textbf{0.0120} & \cellcolor{gray!16}\textbf{0.0236} & \cellcolor{gray!16}\textbf{0.0151} & \cellcolor{gray!16}\textbf{0.0883} & \cellcolor{gray!16}\textbf{0.0507} \\ \midrule
\multicolumn{1}{c|}{\multirow{3}{*}{\textbf{3B}}} & \multicolumn{1}{l|}{\textbf{LETTER}} & 0.0109 & 0.0072 & 0.0151 & 0.0097 & 0.0471 & 0.0236 \\
\multicolumn{1}{c|}{} & \multicolumn{1}{l|}{\textbf{E4SRec}} & 0.0096 & 0.0061 & 0.0129 & 0.0081 & 0.0218 & 0.0103 \\
\multicolumn{1}{c|}{} & \multicolumn{1}{l|}{\cellcolor{gray!16}\textbf{SETRec}} & \cellcolor{gray!16}\textbf{0.0195} & \cellcolor{gray!16}\textbf{0.0123} & \cellcolor{gray!16}\textbf{0.0258} & \cellcolor{gray!16}\textbf{0.0159} & \cellcolor{gray!16}\textbf{0.0964} & \cellcolor{gray!16}\textbf{0.0571} \\ \midrule
\multicolumn{1}{c|}{\multirow{3}{*}{\textbf{7B}}} & \multicolumn{1}{l|}{\textbf{LETTER}} & 0.0099 & 0.0061 & 0.0137 & 0.0081 & 0.0406 & 0.0216 \\
\multicolumn{1}{c|}{} & \multicolumn{1}{l|}{\textbf{E4SRec}} & 0.0088 & 0.0057 & 0.0114 & 0.0072 & 0.0133 & 0.0065 \\
\multicolumn{1}{c|}{} & \multicolumn{1}{l|}{\cellcolor{gray!16}\textbf{SETRec}} & \cellcolor{gray!16}\textbf{0.0194} & \cellcolor{gray!16}\textbf{0.0115} & \cellcolor{gray!16}\textbf{0.0239} & \cellcolor{gray!16}\textbf{0.0140} & \cellcolor{gray!16}\textbf{0.1016} & \cellcolor{gray!16}\textbf{0.0613} \\ \bottomrule
\end{tabular}
}}
\label{tab:scaling_performance}
\end{table}

\subsubsection{\textbf{Effect of Semantic Strength $\bm{\beta}$ (RQ4)}}

\begin{figure}[t]
\vspace{-0.2cm}
\setlength{\abovecaptionskip}{-0.15cm}
\setlength{\belowcaptionskip}{-0.15cm}
  \centering 
  \hspace{-0.105in}
  \subfigure{
  \includegraphics[height=1.4in]{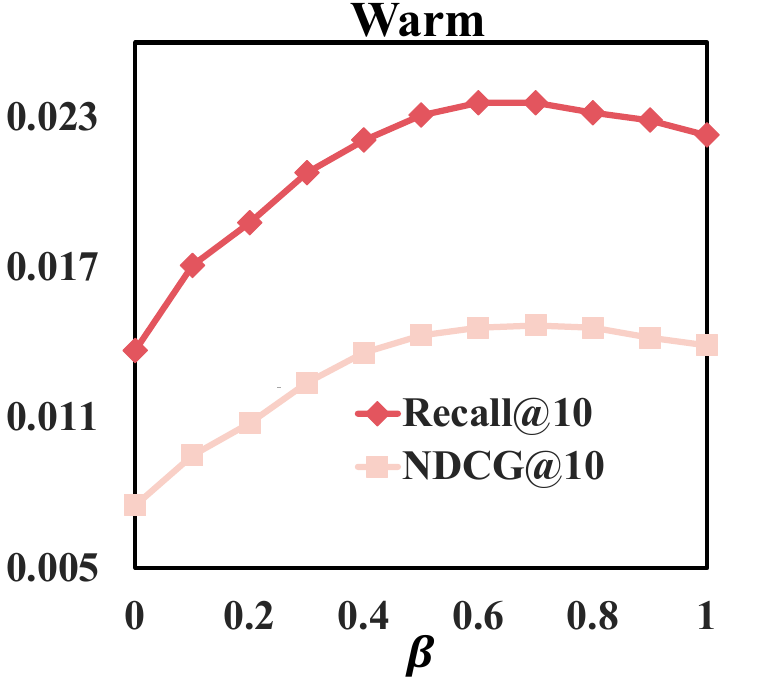}} 
  \hspace{-0.105in}
  \subfigure{    
  \includegraphics[height=1.4in]{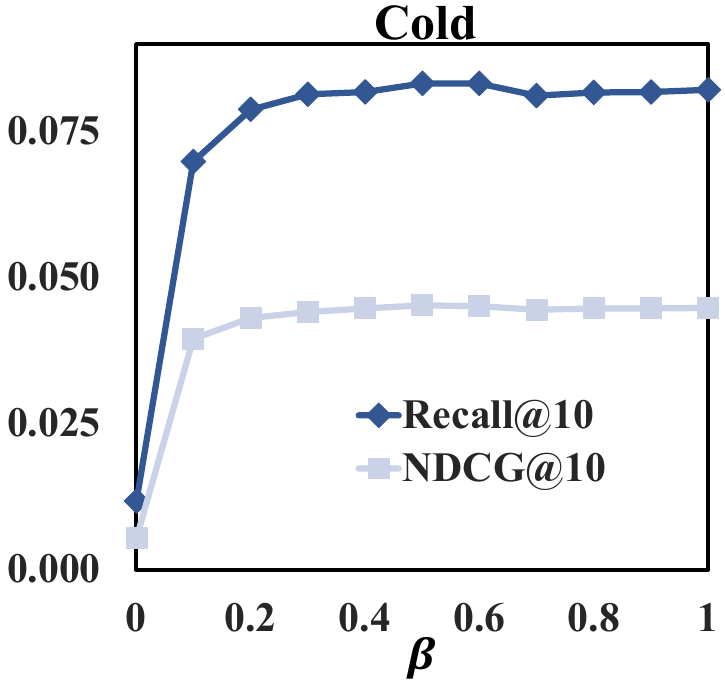}} 
\caption{Performance of SETRec (T5) with different strength of semantics $\beta$ for inference.}
  \label{fig:hp_beta}
\end{figure}

To investigate how semantic information contributes to the performance during inference, we vary $\beta$ from $0$ to $1$, where $\beta=0$ indicates that only CF score is used for ranking, and $\beta=1$ ranks items based solely on semantic scores (Eq. (\ref{eqn:single_logits})).  
From the results in Figure~\ref{fig:hp_beta}, 
we can find that 
1) incorporating semantic information during inference is necessary (inferior performance of $\beta=0$ than $\beta>0$), which facilitates
global ranking over multi-dimensional information for strong generalization ability. 
Notably, 
2) incorporating semantic scores brings more significant improvements on cold items, underscoring the 
semantic information for generalization. 
Moreover, 
3) relying solely on semantics ($\beta=1$), SETRec maintains competitive performance on warm items, which is attributed to the implicit alignment between CF and semantic tokens during training. 

\subsubsection{\textbf{Hyper-parameter Sensitivity (RQ4)}}\label{sec:exp_hyper_param}
\begin{figure}[t]
\vspace{-0.2cm}
\setlength{\abovecaptionskip}{-0.15cm}
\setlength{\belowcaptionskip}{-0cm}
  \centering 
  \hspace{-0.105in}
  \subfigure{
    \includegraphics[height=1.4in]{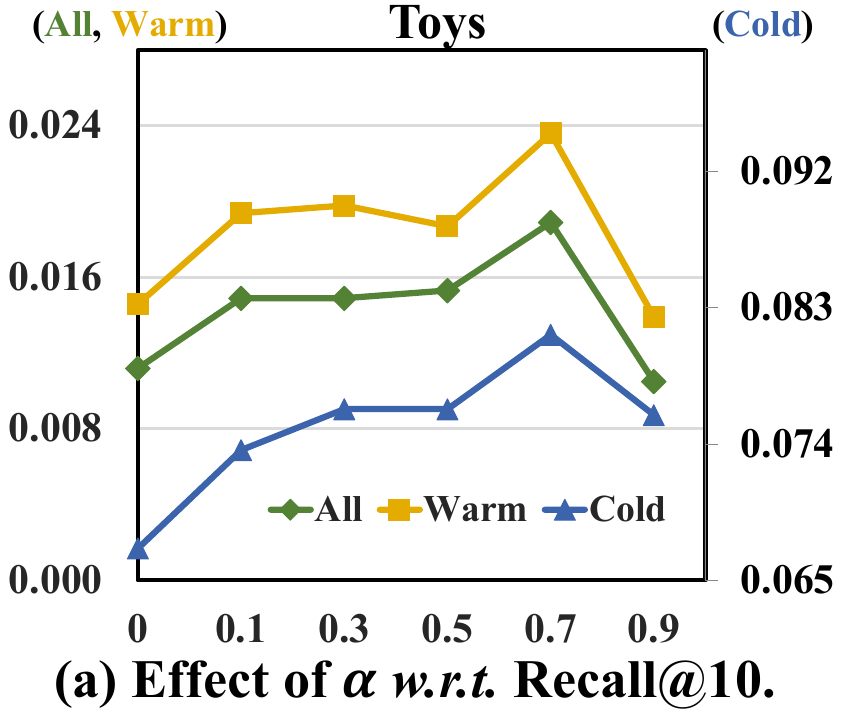}} 
  \subfigure{
    \includegraphics[height=1.4in]{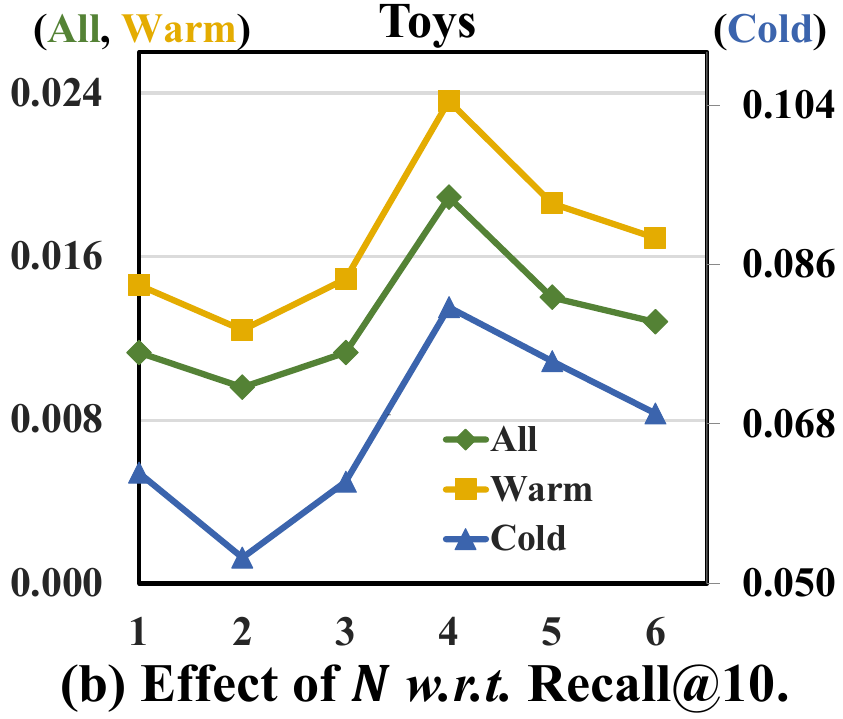}}
\caption{Performance of SETRec (T5) with different strength of AE loss $\alpha$ and different numbers of semantic tokens $N$.}
  \label{fig:hp}
\end{figure}

1) \textbf{Effect of $\bm{\alpha}$}. 
We vary the strength of AE loss $\alpha$ for SETRec training and present the results on Toys in Figure~\ref{fig:hp}(a). 
We can observe that 
the performance is overall improved when $\alpha$ is increased from $0$ to $0.7$, whereas significantly drops on warm items.   
Empirically, we recommend setting $\alpha$ ranging from $0.5$ to $0.7$. 
2) \textbf{Effect of $\bm{N}$}. 
From the results in Figure~\ref{fig:hp}(b), we can find that 
increasing semantic tokens generally improves the performance, 
possibly mitigating the potential information conflicts~\cite{wang2024learnable} and embedding collapse issue~\cite{guoembedding}. 
However, 
blindly increasing the number of semantic tokens might hurt the performance. Because it is non-trivial to recover the category-level preference aligning well with the real-world scenarios~\cite{lin2024disentangled,lin2024temporally}. 


\section{Related Work}\label{sec:related_work}

\noindent$\bullet\quad$\textbf{LLM-based Recommendation.} Harnessing LLMs for recommendations has garnered substantial attention across both academia and industry~\cite{sun2024large,zhang2024gpt4rec,fu2024iisan,shi2024large,zhao2024let,zheng2024harnessing,li2024large,gao2025sprec,li2024survey,xu2025personalized,wang2023diffusion}. 
Existing studies on LLM-based recommendations can be grouped into two research lines. 
1) LLMs for discriminative recommendation, which typically aims to leverage LLMs to assist the conventional recommender models~\cite{wang2024reinforcement,ren2024enhancing,chen2024enhancing,wu2024coral}.                     
This line of work usually involves LLMs in different steps of recommendation pipelines~\cite{lin2023can,zhang2024agentcf,zhang2024generative,wang2024automated} such as feature engineering~\cite{xi2024towards,ren2024representation,xi2024towards} and feature encoder~\cite{chen2024hllm}. 
2) LLMs for generative recommendation, which regards LLMs as recommenders to directly generate items~\cite{liao2024llara,kim2024large,lin2024bridging,wang2024eager,li2024large,lin2024data}. 
To build LLMs for generative recommendation, a key step is item tokenization, where each item is assigned an identifier for LLMs to encode user history and generate the next item. 
In this work, we critically analyze the fundamental principles of identifier design to achieve effective and efficient LLM-based generative recommendation. 

\vspace{2pt}
\noindent$\bullet\quad$\textbf{Identifier for LLM-based Recommendation.} 
Existing identifier designs can be broadly categorized into two types: 
1) token-sequence identifiers represent each item with a discrete token sequence. 
Early studies leverage items' textual information such as titles~\cite{bao2023bi}, descriptions~\cite{cui2022m6}, and tags~\cite{tan2024idgenrec}, aiming to utilize knowledge encoded in LLMs through human vocabulary. 
More recently, utilizing external tokens for identifiers has attracted extensive attention due to its potential to include hierarchical information~\cite{zhu2024cost,zheng2024adapting,wang2024content}. 
Despite the effectiveness and the potential to incorporate both semantic and CF information~\cite{zhang2025collm,wang2024learnable}, token-sequence identifiers suffer from local optima issue and inference inefficiency. 
To improve efficiency, 
2) single-token identifiers are proposed to represent each item with ID or semantic embedding~\cite{li2023e4srec,wang2024rethinking}. 
Nonetheless, existing work neither has poor generalization ability nor fails to capture CF information. 
In this work, we propose a novel set identifier paradigm, which employs a set of order-agnostic CF and semantic tokens.  
Two concurrent studies explore set identifiers for generative retrieval~\cite{zeng2024planning,zhang2024generative}, yet they still preserve token dependencies and heavily rely on autoregressive generation. Differently, our proposed paradigm achieves simultaneous generation without token dependencies, significantly enhancing generation efficiency. 

\section{Conclusion}\label{sec:conclusion}
In this work, we revealed inherent issues of existing identifiers for LLM-based generative recommendation, \ie inadequate information, local optima, and generation inefficiency.    
We then summarized two principles for identifier design, \ie 1) integration of both CF and semantic information, and 2) order-agnostic identifiers. 
Meeting the two principles, we introduced a novel set identifier paradigm, which employs order-agnostic set identifiers to encode user history and generate the set identifier simultaneously. 
To implement this paradigm, we proposed SETRec, which uses CF and semantic tokenizers to obtain a set of CF and semantic tokens. 
To remove token dependencies, we introduced a sparse attention mask for user history encoding and a query-guided generation mechanism for simultaneous generation. 
Empirical results on four datasets across various scenarios demonstrated the effectiveness, efficiency, generalization ability, and scalability of SETRec. 

This work underscores the order agnosticism and multi-dimensional information utilization for identifier design, paving the way for numerous promising avenues for future research. 
%
1) To better align with the pre-training tasks and fully utilize the knowledge within LLMs, it is worth exploring how discrete set identifiers (\ie a set of order-agnostic discrete tokens) perform on generative recommendation. 
2) While SETRec shows strong generalization ability in challenging scenarios such as unpopular item groups, it is worthwhile to apply SETRec for open-ended recommendation with open-domain user behaviors. 

\clearpage

{
\tiny
\bibliographystyle{ACM-Reference-Format}
\balance
\bibliography{bibfile}
}

\newpage

\end{document}